\documentclass[11pt]{article}
\usepackage{amssymb,amsmath,amsfonts}
\usepackage{graphicx}
\usepackage{graphics}
\usepackage{eepic,epsfig}

1.60cm \makeatletter \@addtoreset{equation}{section}

\makeatother

\begin{document}

\title{Vacuum polarization induced by a cylindrical boundary in the cosmic
string spacetime }
\author{E. R. Bezerra de Mello$^{1}$\thanks{%
E-mail: emello@fisica.ufpb.br},\, V. B. Bezerra$^{1}$\thanks{%
E-mail: valdir@fisica.ufpb.br}, \, A. A. Saharian$^{1,2}$\thanks{%
E-mail: saharyan@server.physdep.r.am}, A. S. Tarloyan$^{2}$ \\
\\
\textit{$^{1}$Departamento de F\'{\i}sica, Universidade Federal da Para\'{\i}%
ba}\\
\textit{58.059-970, Caixa Postal 5.008, Jo\~{a}o Pessoa, PB, Brazil}\vspace{0.3cm%
}\\
\textit{$^2$Department of Physics, Yerevan State University,}\\
\textit{375025 Yerevan, Armenia}}
\maketitle

\begin{abstract}
In this paper we investigate the Wightman function, the renormalized vacuum
expectation values of the field square, and the energy-momentum tensor for a
massive scalar field with general curvature coupling inside and outside of a
cylindrical shell in the generalized spacetime of straight cosmic string.
For the general case of Robin boundary condition, by using the generalized
Abel-Plana formula, the vacuum expectation values are presented in the form
of the sum of boundary-free and boundary-induced parts. The asymptotic
behavior of the vacuum expectation values of the field square, energy
density and stresses are investigated in various limiting cases. The
generalization of the results to the exterior region is given for a general
cylindrically symmetric static model of the string core with finite support.
\end{abstract}

\bigskip


\bigskip

\section{Introduction}

Cosmic strings generically arise within the framework of grand unified
theories and could be produced in the early universe as a result of symmetry
braking phase transitions \cite{Kibb80,Vile85}. Although the recent
observational data on the cosmic microwave background radiation have ruled
out cosmic strings as the primary source for primordial density
perturbations, they are still candidates for the generation of a number
interesting physical effects such as the generation of gravitational waves,
high energy cosmic strings, and gamma ray bursts. Recently the cosmic
strings attract a renewed interest partly because a variant of their
formation mechanism is proposed in the framework of brane inflation \cite%
{Sara02,Cope04,Dval04}. In the simplest theoretical model describing the
infinite straight cosmic string the spacetime is locally flat except on the
string where it has a delta shaped curvature tensor. In quantum field theory
the corresponding non-trivial topology leads to non-zero vacuum expectation
values for physical observables. Explicit calculations have been done for
various fields \cite{Hell86}-\cite{Beze06}. The case of quantum fields at
non-zero temperature was also investigated \cite{Cogn94,Line92,Frol95,Line96}%
. Vacuum polarization effects by the cosmic string carying a magnetic flux
are considered in Refs. \cite{Guim95,Srir01,Spin03}. Another type of vacuum
polarization arises when boundaries are present. The imposed boundary
conditions on quantum fields alter the zero-point fluctuations spectrum and
result in additional shifts in the vacuum expectation values of physical
quantities, such as the energy density and stresses. In particular, vacuum
forces arise acting on constraining boundaries. This is the well-known
Casimir effect (for a review see, \cite{Most97}). In this paper we will
study the configuration with both types of sources for the vacuum
polarization, namely, a cylindrical boundary coaxial with a cosmic string
assuming that on the bounding surface the field obeys Robin boundary
condition. For a massive scalar field with an arbitrary curvature coupling
parameter we evaluate the Wightman function and the vacuum expectation
values of the field square and the energy-momentum tensor in both interior
and exterior regions. In addition to describing the physical structure of a
quantum field at a given point, the energy-momentum tensor acts as the
source in the Einstein equations and therefore plays an important role in
modelling a self-consistent dynamics involving the gravitational field. The
vacuum densities for a Robin cylindrical boundary in the Minkowski
background are investigated in \cite{Saha01} (see also \cite{Saha06Tar} for
the case of two coaxial cylindrical surfaces). In Ref. \cite{Khus99} a
cylindrical boundary with Dirichlet boundary condition is introduced in the
bulk of the cosmic string as an intermediate stage for the calculation of
the ground state energy of a massive scalar field in (2+1)-dimensions.

We have organized the paper as follows. The next section is devoted to the
evaluation of the Wightman function for a massive scalar field in a
generalized cosmic string spacetime in both interior and exterior regions of
a coaxial cylindrical boundary. In section \ref{sec:noboundary} we present
simple formulae for the vacuum expectation values of the field square and
the energy-momentum tensor in the corresponding boundary-free geometry. By
using the formula for the Wightman function, in section \ref{sec:inside} we
evaluate the vacuum expectation values of the field square and the
energy-momentum tensor inside a cylindrical boundary. Various limiting cases
are investigated. In section \ref{sec:outside} we consider the corresponding
quantities for the outside region. The generalization of the results for the
exterior region in the case of a general cylindrically symmetric static
model of the string core with finite support is given in section \ref%
{sec:finitethick}. Finally, the results are summarized and discussed in
section \ref{sec:Conc}.

\section{Wightman function}

\label{sec:WightFunc}

\subsection{Bulk and boundary geometries}

In this paper we consider a scalar field $\varphi $ propagating on the $(D+1)
$-dimensional background spacetime with a conical-type singularity described
by the line-element%
\begin{equation}
ds^{2}=g_{ik}dx^{i}dx^{k}=dt^{2}-dr^{2}-r^{2}d\phi
^{2}-\sum_{i=1}^{N}dz_{i}{}^{2},  \label{ds21}
\end{equation}%
with the cylindrical coordinates $(x^{1},x^{2},\ldots ,x^{D})=(r,\phi
,z_{1},\ldots ,z_{N})$, where $N=D-2$, $r\geqslant 0$, $0\leqslant \phi
\leqslant \phi _{0}$, $-\infty <z_{i}<+\infty $ and the spatial points $%
(r,\phi ,z_{1},\ldots ,z_{N})$ and $(r,\phi +\phi _{0},z_{1},\ldots ,z_{N})$
are to be identified. In the standard $D=3$ cosmic string case the planar
angle deficit is related to the mass per unit length of the string $\mu $ by
$2\pi -\phi _{0}=8\pi G\mu $, where $G$ is the Newton gravitational
constant. It is interesting to note that the effective metric produced in
superfluid $^{3}\mathrm{He-A}$ by a radial disgyration is described by the $%
D=3$ line element (\ref{ds21}) with the negative angle deficit, that is $%
\phi _{0}>2\pi $, which corresponds to the negative mass of the topological
object \cite{Volo98}.

For a free massive field with curvature coupling parameter $\xi $ the field
equation has the form
\begin{equation}
\left( \nabla ^{i}\nabla _{i}+m^{2}+\xi R\right) \varphi (x)=0,
\label{fieldeq}
\end{equation}%
where $\nabla _{i}$ is the covariant derivative operator and $R$ is the
scalar curvature for the background spacetime. The values of the curvature
coupling parameter $\xi =0$ and $\xi =\xi _{D}\equiv (D-1)/4D$ correspond to
the most important special cases of minimally and conformally coupled
scalars, respectively. We assume that the field obeys Robin boundary
condition on the cylindrical surface with radius $a$, coaxial with the
string:
\begin{equation}
\left( A+B\frac{\partial }{\partial r}\right) \varphi =0,\quad r=a.
\label{Dirbc}
\end{equation}%
Of course, all results in what follows will depend on the ratio of the
coefficients in this boundary condition. However, to keep the transition to
Dirichlet and Neumann cases transparent, we use the form (\ref{Dirbc}). In
this section we are interested in the corresponding positive frequency
Wightman function in the regions inside and outside of the cylindrical
surface due to the fact that the vacuum expectation values (VEVs) of the
field square and the energy-momentum tensor are expressed in terms of this
function. In addition, the response of a particle detector in an arbitrary
state of motion is determined by this function (see, for instance, \cite%
{Birr82}). By expanding the field operator in terms of a complete set of
eigenfunctions $\{\varphi _{\mathbf{\alpha }}(x),\varphi _{\mathbf{\alpha }%
}^{\ast }(x)\}$ satisfying the boundary condition and using the standard
commutation relations, the Wightman function is presented as the mode-sum
\begin{equation}
\langle 0|\varphi (x)\varphi (x^{\prime })|0\rangle =\sum_{\mathbf{\alpha }%
}\varphi _{\mathbf{\alpha }}(x)\varphi _{\mathbf{\alpha }}^{\ast }(x),
\label{vevWf}
\end{equation}%
where $\alpha $ is a collective notation for the quantum numbers specifying
the solution and $|0\rangle $ is the amplitude for the corresponding vacuum
state. The form of the eigenfunctions is different for the regions inside
and outside the cylindrical shell and we will consider these regions
separately.

\subsection{Wightman function in the region inside the shell}

In the region inside the cylindrical surface the eigenfunctions satisfying
the periodicity condition on $\phi =\phi _{0}$ are specified by the set of
quantum numbers $\alpha =(n,\gamma ,\mathbf{k})$, $n=0,\pm 1,\pm 2,\cdots $,
$\mathbf{k}=(k_{1},\ldots ,k_{N})$, $-\infty <k_{j}<\infty $, and have the
form
\begin{eqnarray}
\varphi _{\alpha }(x) &=&\beta _{\alpha }J_{q\left\vert n\right\vert
}(\gamma r)\exp \left( iqn\phi +i\mathbf{kr}_{\parallel }-i\omega t\right) ,
\label{eigfunccirc} \\
\omega &=&\sqrt{\gamma ^{2}+k^{2}+m^{2}},\quad q=2\pi /\phi _{0},  \label{qu}
\end{eqnarray}%
where $\mathbf{r}_{\parallel }=(z_{1},\ldots ,z_{N})$ and $J_{l}(z)$ is the
Bessel function. The eigenvalues for the quantum number $\gamma $ are
quantized by the boundary condition (\ref{Dirbc}) on the cylindrical surface
$r=a$. From this condition it follows that for a given $n$ the possible
values of $\gamma $ are determined by the relation
\begin{equation}
\gamma =\lambda _{n,j}/a,\quad j=1,2,\cdots ,  \label{ganval}
\end{equation}%
where $\lambda _{n,j}$ are the positive zeros of the function $\bar{J}%
_{q|n|}(z)$, $\bar{J}_{q|n|}(\lambda _{n,j})=0$, arranged in ascending
order, $\lambda _{n,j}<\lambda _{n,j+1}$, $n=0,1,2,\ldots $. Here and in
what follows, for a given function $f(z)$, we use the notation%
\begin{equation}
\bar{f}(z)=Af(z)+\left( B/a\right) zf^{\prime }(z).  \label{fbar}
\end{equation}%
It is well-known that for real $A$ and $B$ all zeros of the function $\bar{J}%
_{q|n|}(z)$ are simple and real, except the case $Aa/B<-q|n|$ when there are
two purely imaginary zeros. In the following we will assume the values of $%
Aa/B$ for which all zeros are real.

The coefficient $\beta _{\alpha }$ in (\ref{eigfunccirc}) is determined from
the normalization condition based on the standard Klein-Gordon scalar
product with the integration over the region inside the cylindrical surface
and is equal to
\begin{equation}
\beta _{\alpha }^{2}=\frac{\lambda _{n,j}T_{qn}(\lambda _{n,j})}{(2\pi
)^{N}\omega \phi _{0}a^{2}},  \label{betalf}
\end{equation}%
where we have introduced the notation%
\begin{equation}
T_{qn}(z)=z\left[ \left( z^{2}-q^{2}n^{2}\right)
J_{q|n|}^{2}(z)+z^{2}J_{q|n|}^{\prime 2}(z)\right] ^{-1}.  \label{Tqn}
\end{equation}%
Substituting the eigenfunctions (\ref{eigfunccirc}) into the mode-sum
formula (\ref{vevWf}) with a set of quantum numbers $\alpha =(nj\mathbf{k})$%
, for the positive frequency Wightman function one finds
\begin{eqnarray}
\langle 0|\varphi (x)\varphi (x^{\prime })|0\rangle &=&2\int d^{N}\mathbf{k}%
\,e^{i\mathbf{k}(\mathbf{r}_{\parallel }-\mathbf{r}_{\parallel }^{\prime })}%
\sideset{}{'}{\sum}_{n=0}^{\infty }\cos [qn(\phi -\phi ^{\prime })]  \notag
\\
&&\times \sum_{j=1}^{\infty }\beta _{\alpha }^{2}J_{qn}(\gamma
r)J_{qn}(\gamma r^{\prime })e^{-i\omega (t-t^{\prime })}|_{\gamma =\lambda
_{n,j}/a},  \label{Wf1}
\end{eqnarray}%
where the prime means that the summand with $n=0$ should be taken with the
weight 1/2. As we do not know the explicit expressions for the eigenvalues $%
\lambda _{n,j}$ as functions on $n$ and $j$, and the summands in the series
over $j$ are strongly oscillating functions for large values of $j$, this
formula is not convenient for the further evaluation of the VEVs of the
field square and the energy-momentum tensor. In addition, the expression on
the right of (\ref{Wf1}) is divergent in the coincidence limit and some
renormalization procedure is needed to extract finite result for the VEVs of
the field square and the energy-momentum tensor. To obtain an alternative
form for the Wightman function we will apply to the sum over $j$ a variant
of the generalized Abel-Plana summation formula \cite{Saha87}
\begin{eqnarray}
\sum_{j=1}^{\infty }T_{qn}(\lambda _{n,j})f(\lambda _{n,j}) &=&\frac{1}{2}%
\int_{0}^{\infty }dz\,f(z)-\frac{1}{2\pi }\int_{0}^{\infty }dz\,\frac{\bar{K}%
_{qn}(z)}{\bar{I}_{qn}(z)}  \notag \\
&&\times \left[ e^{-qn\pi i}f(ze^{\frac{\pi i}{2}})+e^{qn\pi i}f(ze^{-\frac{%
\pi i}{2}})\right] ,  \label{sumform1AP}
\end{eqnarray}%
where $I_{l}(z)$ and $K_{l}(z)$ are the modified Bessel functions. This
formula is valid for functions $f(z)$ analytic in the right half-plane of
the complex variable $z=x+iy$ and satisfying the condition $|f(z)|<\epsilon
(x)e^{c_{1}|y|}$, $c_{1}<2$, for $\left\vert z\right\vert \rightarrow \infty
$ and the condition $f(z)=o(z^{2q\left\vert n\right\vert -1})$ for $%
z\rightarrow 0$, where $\epsilon (x)\rightarrow 0$ for $x\rightarrow \infty $%
. By taking in formula (\ref{sumform1AP}) $qn=1/2$ and $B=0$, as a special
case, we obtain the Abel-Plana formula.

To evaluate the sum over $j$ in (\ref{Wf1}) as a function $f(z)$ we choose
\begin{equation}
f(z)=\frac{zJ_{qn}(zr/a)J_{qn}(zr^{\prime }/a)}{\sqrt{k^{2}+m^{2}+z^{2}/a^{2}%
}}\exp \left[ -i\sqrt{k^{2}+m^{2}+z^{2}/a^{2}}(t-t^{\prime })\right] .
\label{ftosum}
\end{equation}%
Using the asymptotic formulae of the Bessel functions for large values of
the argument when $n$ is fixed (see, e.g., \cite{hand}), we can see that for
the function $f(z)$ given in (\ref{ftosum}), the condition to formula (\ref%
{sumform1AP}) to be satisfied is $r+r^{\prime }+|t-t^{\prime }|<2a$. In
particular, this is the case in the coincidence limit $t=t^{\prime }$ for
the region under consideration, $r,r^{\prime }<a$. Formula (\ref{sumform1AP}%
) allows to present the Wightman function in the form%
\begin{equation}
\langle 0|\varphi (x)\varphi (x^{\prime })|0\rangle =\langle 0_{s}|\varphi
(x)\varphi (x^{\prime })|0_{s}\rangle +\langle \varphi (x)\varphi (x^{\prime
})\rangle _{a},  \label{Wf2}
\end{equation}%
where%
\begin{eqnarray}
\langle 0_{s}|\varphi (x)\varphi (x^{\prime })|0_{s}\rangle  &=&\frac{1}{%
\phi _{0}}\int \frac{d^{N}\mathbf{k}}{(2\pi )^{N}}e^{i\mathbf{k}(\mathbf{r}%
_{\parallel }-\mathbf{r}_{\parallel }^{\prime })}\int_{0}^{\infty }dz\frac{%
ze^{-i(t-t^{\prime })\sqrt{z^{2}+k^{2}+m^{2}}}}{\sqrt{z^{2}+k^{2}+m^{2}}}
\notag \\
&&\times \sideset{}{'}{\sum}_{n=0}^{\infty }\cos [qn(\phi -\phi ^{\prime
})]J_{qn}(zr)J_{qn}(zr^{\prime }),  \label{Wf00}
\end{eqnarray}%
and%
\begin{eqnarray}
\langle \varphi (x)\varphi (x^{\prime })\rangle _{a} &=&-\frac{2}{\pi \phi
_{0}}\int \frac{d^{N}\mathbf{k}}{(2\pi )^{N}}e^{i\mathbf{k}(\mathbf{r}%
_{\parallel }-\mathbf{r}_{\parallel }^{\prime })}\int_{\sqrt{k^{2}+m^{2}}%
}^{\infty }dz\frac{z\cosh \left[ (t-t^{\prime })\sqrt{z^{2}-k^{2}-m^{2}}%
\right] }{\sqrt{z^{2}-k^{2}-m^{2}}}  \notag \\
&&\times \sideset{}{'}{\sum}_{n=0}^{\infty }\cos [qn(\phi -\phi ^{\prime
})]I_{qn}(zr)I_{qn}(zr^{\prime })\frac{\bar{K}_{qn}(za)}{\bar{I}_{qn}(za)}.
\label{Wfa0}
\end{eqnarray}%
In the limit $a\rightarrow \infty $ for fixed $r,r^{\prime }$, the term $%
\langle \varphi (x)\varphi (x^{\prime })\rangle _{a}$ vanishes and, hence,
the term $\langle 0_{s}|\varphi (x)\varphi (x^{\prime })|0_{s}\rangle $ is
the Wightman function for the geometry of a cosmic string without a
cylindrical boundary with the corresponding vacuum state $|0_{s}\rangle $.
Consequently, the term $\langle \varphi (x)\varphi (x^{\prime })\rangle _{a}$
is induced by the presence of the cylindrical boundary. Hence, the
application of the generalized Abel-Plana formula allows us to extract from
the Wightman function the part due to the string without a cylindrical
boundary. For points away from the cylindrical surface the additional part
induced by this surface, formula (\ref{Wfa0}), is finite in the coincidence
limit and the renormalization is needed only for the part coming from (\ref%
{Wf00}).

\subsection{Wightman function in the exterior region}

\label{subsec:exterior}

Now we turn to the region outside the cylindrical shell, $r>a$. In general,
the coefficients in Robin boundary condition (\ref{Dirbc}) for this region
can be different from those for the interior region. However, in order to
not complicate the formulae we use the same notations. The corresponding
eigenfunctions satisfying boundary conditions (\ref{Dirbc}) are obtained
from (\ref{eigfunccirc}) by the replacement
\begin{equation}
J_{qn}(\gamma r)\rightarrow g_{qn}(\gamma r,\gamma a)\equiv J_{qn}(\gamma r)%
\bar{Y}_{qn}(\gamma a)-\bar{J}_{qn}(\gamma a)Y_{qn}(\gamma r),
\label{replace}
\end{equation}%
where $Y_{qn}(z)$ is the Neumann function. Now the spectrum for
the quantum number $\gamma $ is continuous. To determine the
corresponding normalization coefficient $\beta _{\alpha }$, we
note that as the normalization integral diverges in the limit
$\gamma =\gamma ^{\prime }$, the main contribution into the
integral over radial coordinate comes from the large values of $r$
when the Bessel functions can be replaced by their asymptotics for
large arguments. The resulting integral is elementary and for the
normalization
coefficient in the region $r>a$ one finds%
\begin{equation}
\beta _{\alpha }^{2}=\frac{(2\pi )^{2-D}\gamma }{2\phi _{0}\omega \left[
\bar{J}_{qn}^{2}(\gamma a)+\bar{Y}_{qn}^{2}(\gamma a)\right] }.
\label{norcoefext}
\end{equation}%
Substituting the corresponding eigenfunctions into the mode-sum formula (\ref%
{vevWf}), the positive frequency Whightman function can be written in the
form
\begin{eqnarray}
\langle 0|\varphi (x)\varphi (x^{\prime })|0\rangle  &=&\frac{1}{\phi _{0}}%
\int \frac{d^{N}{\mathbf{k}}}{(2\pi )^{N}}e^{i{\mathbf{k}}({\mathbf{r}}%
_{\parallel }-{\mathbf{r}}_{\parallel }^{\prime })}\sideset{}{'}{\sum}%
_{n=0}^{\infty }\cos [qn(\phi -\phi ^{\prime })]  \notag \\
&&\times \int_{0}^{\infty }d\gamma \frac{\gamma g_{qn}(\gamma r,\gamma
a)g_{qn}(\gamma r^{\prime },\gamma a)}{\bar{J}_{qn}^{2}(\gamma a)+\bar{Y}%
_{qn}^{2}(\gamma a)}\frac{\exp \left[ i(t^{\prime }-t)\sqrt{\gamma ^{2}+k^{2}%
}\right] }{\sqrt{\gamma ^{2}+k^{2}}}.  \label{Wfext0}
\end{eqnarray}%
To find the part in the Wightman function induced by the presence of the
cylindrical shell we subtract from (\ref{Wfext0}) the corresponding function
for the geometry of a cosmic string without a cylindrical shell, given by (%
\ref{Wf00}). In order to evaluate the corresponding difference we use the
relation
\begin{equation}
\frac{g_{qn}(\gamma r,\gamma a)g_{qn}(\gamma r^{\prime },\gamma a)}{\bar{J}%
_{qn}^{2}(\gamma a)+\bar{Y}_{qn}^{2}(\gamma a)}-J_{qn}(\gamma
r)J_{qn}(\gamma r^{\prime })=-\frac{1}{2}\sum_{s=1}^{2}\frac{\bar{J}%
_{qn}(\gamma a)}{H_{qn}^{(s)}(\gamma a)}H_{qn}^{(s)}(\gamma
r)H_{qn}^{(s)}(\gamma r^{\prime }),  \label{relext}
\end{equation}%
where $H_{qn}^{(s)}(z)$, $s=1,2$ are the Hankel functions. This allows to
present the Wightman function in the form (\ref{Wf2}) with the cylindrical
shell induced part given by
\begin{eqnarray}
\langle \varphi (x)\varphi (x^{\prime })\rangle _{a} &=&-\frac{1}{2\phi _{0}}%
\int \frac{d^{N}{\mathbf{k}}}{(2\pi )^{N}}e^{i{\mathbf{k}}({\mathbf{r}}%
_{\parallel }-{\mathbf{r}}_{\parallel }^{\prime })}\sideset{}{'}{\sum}%
_{n=0}^{\infty }\cos [qn(\phi -\phi ^{\prime
})]\sum_{s=1}^{2}\int_{0}^{\infty }d\gamma \gamma   \notag \\
&&\times \frac{H_{qn}^{(s)}(\gamma r)H_{qn}^{(s)}(\gamma r^{\prime })}{\sqrt{%
\gamma ^{2}+k^{2}}}\frac{\bar{J}_{qn}(\gamma a)}{H_{qn}^{(s)}(\gamma a)}\exp %
\left[ i(t^{\prime }-t)\sqrt{\gamma ^{2}+k^{2}}\right] .  \label{Wfext2}
\end{eqnarray}%
On the complex plane $\gamma $, we rotate the integration contour on the
right of this formula by the angle $\pi /2$ for $s=1$ and by the angle $-\pi
/2$ for $s=2$. The integrals over the segments $(0,ik)$ and $(0,-ik)$ cancel
out and after introducing the modified Bessel functions we obtain
\begin{eqnarray}
\langle \varphi (x)\varphi (x^{\prime })\rangle _{a} &=&-\frac{2}{\pi \phi
_{0}}\int \frac{d^{N}\mathbf{k}}{(2\pi )^{N}}e^{i\mathbf{k}(\mathbf{r}%
_{\parallel }-\mathbf{r}_{\parallel }^{\prime })}\int_{k}^{\infty }dz\frac{%
z\cosh \left[ (t-t^{\prime })\sqrt{z^{2}-k^{2}}\right] }{\sqrt{z^{2}-k^{2}}}
\notag \\
&&\times \sideset{}{'}{\sum}_{n=0}^{\infty }\cos [qn(\phi -\phi ^{\prime
})]K_{qn}(zr)K_{qn}(zr^{\prime })\frac{\bar{I}_{qn}(za)}{\bar{K}_{qn}(za)}.
\label{Wfa0ext}
\end{eqnarray}%
As we see the boundary induced part of the Wightman function for the
exterior region is obtained from the corresponding part in the interior
region by the replacements $I\rightleftarrows K$. Note that in formulae (\ref%
{Wf00}), (\ref{Wfa0}) and (\ref{Wfa0ext}) the integration over the
directions of the vector $\mathbf{k}$ can be done using the formula%
\begin{equation}
\int \frac{d^{N}{\mathbf{k}}}{(2\pi )^{N}}e^{i{\mathbf{kr}}%
}F(k)=\int_{0}^{\infty }\frac{dk}{(2\pi )^{N/2}}k^{N-1}F(k)\frac{J_{N/2-1}(k|%
\mathbf{x}|)}{(k|\mathbf{x}|)^{N/2-1}},  \label{intformoverk}
\end{equation}%
for a given function $F(k)$.

\section{Vacuum expectation values for a string without a cylindrical
boundary}

\label{sec:noboundary}

In this section we consider the geometry of a string without a cylindrical
boundary. The one-loop quantum effects of the scalar field in this geometry
have been considered in a large number of papers. The VEV of the
energy-momentum tensor for a conformally coupled $D=3$ massless scalar field
was evaluated in Ref. \cite{Hell86}. The case of an arbitrary curvature
coupling is considered in Refs. \cite{Frol87,Dowk87,Smit89}. The integral
representation for the Green's function for a massive scalar field is
considered in Refs. \cite{Line87,Shir92,Guim94,More95}. The corresponding
local zeta functions are discussed in Refs. \cite{Cogn94,Iell97}. However,
to our knowledge, no closed formulae have been given in the literature for
the VEVs of the field square and the energy-momentum tensor in the case of a
massive field and in an arbitrary number of dimensions. The mass corrections
in the limit $mr\ll 1$ have been considered in Refs. \cite{More95,Iell97}.
Below we will derive simple exact formulae for the VEVs of both field square
and the energy-momentum tensor assuming that the parameter $q$ is an
integer. These formulae will also shed light on the qualitative behavior of
the VEVs in the case of non-integer $q$. The Wightman function for the
geometry of a string without cylindrical boundary is given by formula (\ref%
{Wf00}). For a field with the mass $m$ and in the case of $N$ parallel
dimensions along the string we denote this function as $G_{(N)}^{+}(x,x^{%
\prime },m)$. The following recurrence relation is a simple consequence of
formula (\ref{Wf00}):%
\begin{equation}
G_{(N)}^{+}(x,x^{\prime },m)=\frac{1}{\pi }\int_{0}^{\infty }dk_{N}\cos %
\left[ k_{N}(z_{N}-z_{N}^{\prime })\right] G_{(N-1)}^{+}(x,x^{\prime },\sqrt{%
k_{N}^{2}+m^{2}}).  \label{RecW}
\end{equation}%
Of course, this formula takes place for any other two-point function (see
Ref. \cite{Guim94} for the case of Green function). After the integration
over the angular part of the wave vector $\mathbf{k}$ using formula (\ref%
{intformoverk}), the integral over $k$ is evaluated on the base of formulae
given in Ref. \cite{Prud86}. Assuming that $|\mathbf{r}_{\parallel }-\mathbf{%
r}_{\parallel }^{\prime }|>|t-t^{\prime }|$, one obtains the following
expression%
\begin{eqnarray}
\langle 0_{s}|\varphi (x)\varphi (x^{\prime })|0_{s}\rangle  &=&\frac{%
v^{1-\nu }}{2^{\nu -1}\pi ^{\nu }\phi _{0}}\int_{0}^{\infty }dz\,z\left(
z^{2}+m^{2}\right) ^{\frac{\nu -1}{2}}K_{\nu -1}(v\sqrt{z^{2}+m^{2}})  \notag
\\
&&\times \sideset{}{'}{\sum}_{n=0}^{\infty }\cos [qn(\phi -\phi ^{\prime
})]J_{qn}(zr)J_{qn}(zr^{\prime }),  \label{Wfpurestring}
\end{eqnarray}%
where we have introduced the notations%
\begin{equation}
\nu =\frac{D-1}{2},\;v=|(\mathbf{r}_{\parallel }-\mathbf{r}_{\parallel
}^{\prime })^{2}-(t-t^{\prime })^{2}|^{1/2}.  \label{ve}
\end{equation}%
The corresponding formula in the case $|\mathbf{r}_{\parallel }-\mathbf{r}%
_{\parallel }^{\prime }|<|t-t^{\prime }|$ is obtained from (\ref%
{Wfpurestring}) by the replacement $K_{\nu -1}\rightarrow (\pi /2)e^{i\pi
(\nu -1/2)}H_{\nu -1}^{(1)}$. In special cases, formula (\ref{Wfpurestring})
agrees with the results form Refs. \cite{Smit89,Shir92}. The analog formula
for the Green's function in the case of more general background of the
spinning cone and for a twisted scalar field is given in \cite{More95}. For
integer values of the parameter $q$ the expression for the Wightman function
can be further simplified by using the formula \cite{Prud86} (see also Ref.
\cite{Davi88})%
\begin{equation}
\sideset{}{'}{\sum}_{n=0}^{\infty }\cos [qn(\phi -\phi ^{\prime
})]J_{qn}(zr)J_{qn}(zr^{\prime })=\frac{1}{2q}\sum_{l=0}^{q-1}J_{0}\left(
zu_{l}\right) ,  \label{formn1}
\end{equation}%
where
\begin{equation}
u_{l}=r^{2}+r^{\prime 2}-2rr^{\prime }\cos (l\phi _{0}+\phi -\phi ^{\prime
}).  \label{njpm}
\end{equation}%
By using (\ref{formn1}) in Eq. (\ref{Wfpurestring}) and evaluating the
integral over $z$ with the help of a formula from \cite{Prud86}, we arrive
at the following expression%
\begin{equation}
\langle 0_{s}|\varphi (x)\varphi (x^{\prime })|0_{s}\rangle =\frac{m^{\frac{%
D-1}{2}}}{(2\pi )^{\frac{D+1}{2}}}\sum_{l=0}^{q-1}\frac{K_{\nu }(m\sqrt{%
u_{l}^{2}+v^{2}})}{\left( u_{l}^{2}+v^{2}\right) ^{\nu /2}},  \label{Wf02n}
\end{equation}%
where $\nu $ and $v$ are defined by formulae (\ref{ve}). The $l=0$ term in
formula (\ref{Wf02n}) coincides with the Wightman function of the Minkowski
space without boundaries, and the Wightman function for the geometry of the
string is a sum of $q$ images of the Minkowski spacetime functions. We could
obtain this result directly by using the method of images (see also \cite%
{Smit89,Sour92} for $D=3$ and $D=2$ massless cases).

The VEV of the field square is formally given by taking the coincidence
limit $x^{\prime }\rightarrow x$ of the Wightman function. However, this
procedure leads to a divergent result. In order to obtain a finite and well
defined result, we apply the renormalization procedure which corresponds to
the subtraction from the Wightman function in the geometry of the cosmic
string the corresponding function in the background Minkowski spacetime. So
the renormalized VEV\ of the field square reads%
\begin{eqnarray}
\langle \varphi ^{2}\rangle _{\mathrm{ren}}^{(s)} &=&\langle 0_{s}|\varphi
^{2}|0_{s}\rangle -\langle 0_{M}|\varphi ^{2}|0_{M}\rangle   \notag \\
&=&\frac{m^{D-1}}{(2\pi )^{\frac{D+1}{2}}}\sum_{l=1}^{q-1}\frac{K_{\nu
}(2mry_{l})}{(2mry_{l})^{\nu }},  \label{phi2w}
\end{eqnarray}%
where we used the notation%
\begin{equation}
\;y_{l}=\sin (\pi l/q).  \label{nj}
\end{equation}%
From (\ref{phi2w}) it follows that the renormalized VEV of the field square
is positive everywhere. In the limit $mr\gg 1/\sin (\pi /q)$, the main
contribution to the VEV (\ref{Wf02n}) comes from the terms with $l=1$ and $%
l=q-1$ and to the leading order one finds%
\begin{equation}
\langle \varphi ^{2}\rangle _{\mathrm{ren}}^{(s)}\approx \frac{m^{D/2-1}}{%
(4\pi r)^{D/2}}\frac{e^{-2mr\sin (\pi /q)}}{\sin ^{D/2}(\pi /q)},
\label{phi20largem}
\end{equation}%
with exponentially suppressed VEV.

In the case of a massless scalar field we obtain from (\ref{phi2w}) the
following result%
\begin{equation}
\langle \varphi ^{2}\rangle _{\mathrm{ren}}^{(s)}=\frac{\Gamma ((D-1)/2)}{%
(4\pi )^{\frac{D+1}{2}}r^{D-1}}\sum_{l=1}^{q-1}y_{l}^{1-D}.
\label{phi2wmassless}
\end{equation}%
For the case $D=2$ this agree with the result in Ref. \cite{Sour92}. From (%
\ref{phi2w}) it follows that in the case of a massive field the expression
on the right of this formula determines the leading term in the asymptotic
expansion of the VEV\ of the field square near the string. For odd values of
$D$ the summation on the right of formula (\ref{phi2wmassless}) can be done
by making use of the formulae%
\begin{equation}
\mathcal{I}_{n+2}(x)=\frac{\mathcal{I}_{n}^{\prime \prime }(x)+n^{2}\mathcal{%
I}_{n}(x)}{n(n+1)},\quad \mathcal{I}_{2}(x)=q^{2}\sin ^{-2}(qx)-\sin ^{-2}x,
\label{In+2}
\end{equation}%
for the sum%
\begin{equation}
\mathcal{I}_{n}(x)=\sum_{l=1}^{q-1}\sin ^{-n}\left( x+l\pi /q\right) .
\label{Inx}
\end{equation}%
In particular, one has $\mathcal{I}_{2}(0)=(q^{2}-1)/3$ and in the case $D=3$
we obtain the well-known result for the VEV of the field square in the
massless case \cite{Line87,Smit89}. For $D=5$, by using recurrence relation (%
\ref{In+2}) for the evaluation of $\mathcal{I}_{4}(x)$, we find $\mathcal{I}%
_{4}(0)=(q^{2}-1)(q^{2}+11)/45$ and%
\begin{equation}
\langle \varphi ^{2}\rangle _{\mathrm{ren}}^{(s)}=\frac{(q^{2}-1)(q^{2}+11)}{%
2880\pi ^{3}r^{4}},\;D=5.  \label{phi2D5}
\end{equation}%
It is worth call attention to the fact that the resulting expressions are
analytic functions of $q$ and by analytic continuation they are valid for
all values of $q$.

\begin{figure}[tbph]
\begin{center}
\begin{tabular}{cc}
\epsfig{figure=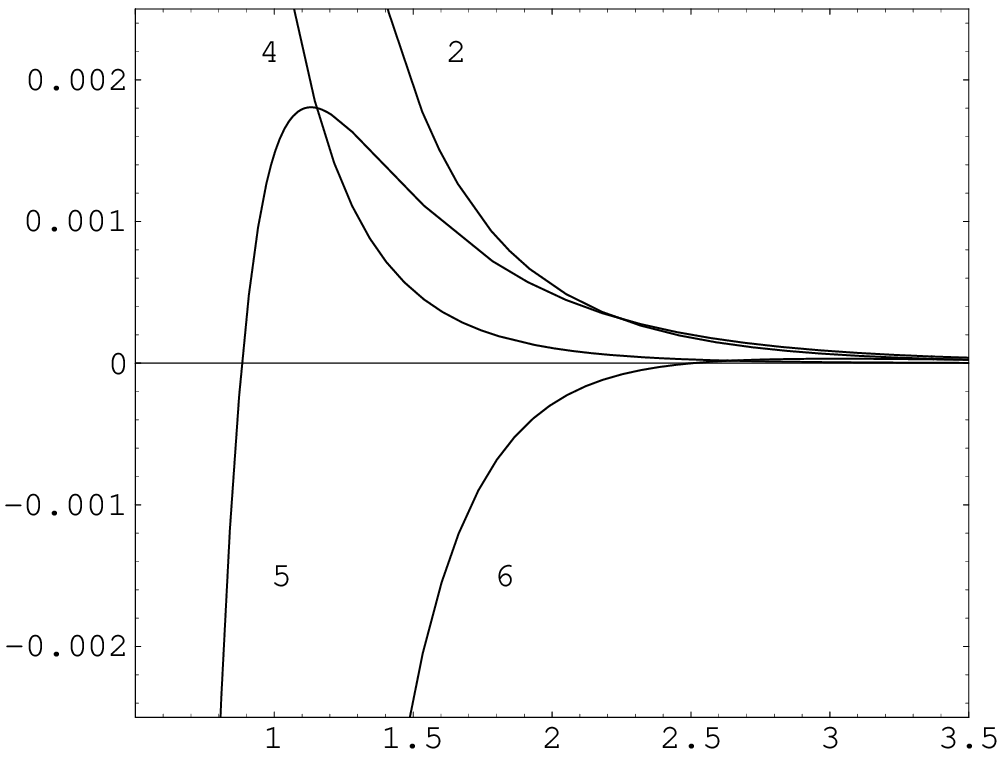,width=7.cm,height=5.5cm} & \quad %
\epsfig{figure=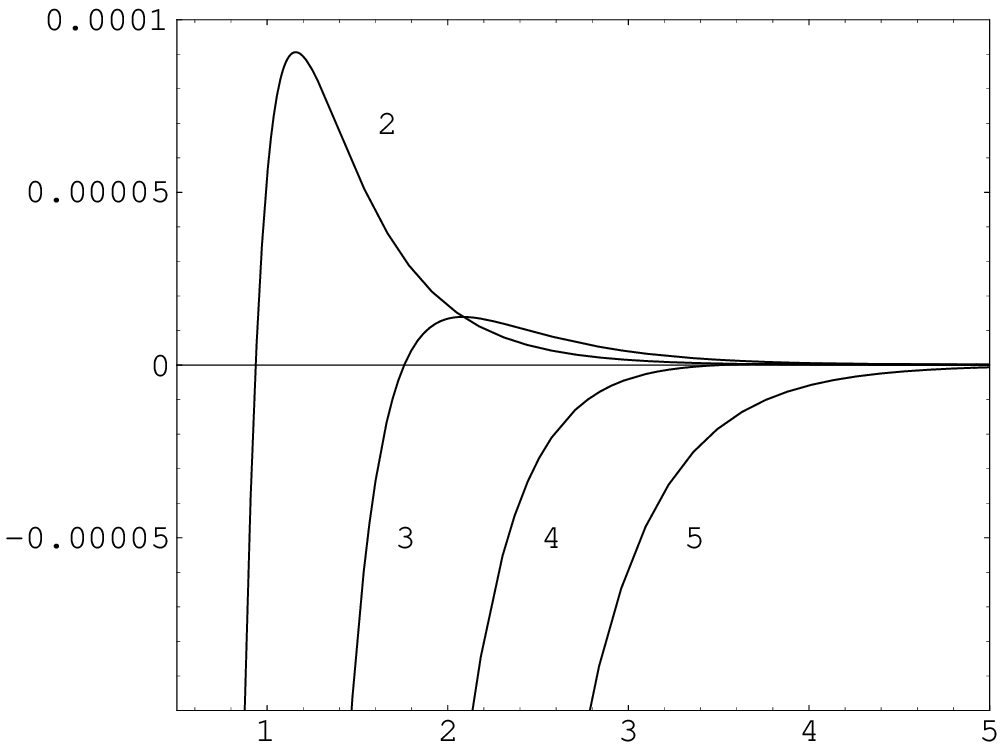,width=7.cm,height=5.5cm}%
\end{tabular}%
\end{center}
\caption{Vacuum energy density divided by $m^{D+1}$, $\langle
T_{0}^{0}\rangle _{\mathrm{ren}}^{(s)}/m^{D+1}$, for minimally (left panel)
and conformally (right panel) coupled $D=3$ scalar fields as a function on $%
mr $ for various values of the parameter $q$ (the numbers near the curves).}
\label{fig1}
\end{figure}

Now we turn to the VEVs of the energy-momentum tensor in the cosmic string
spacetime without boundaries. These VEVs can be evaluated by making use of
the formula
\begin{equation}
\langle 0|T_{ik}(x)|0\rangle =\lim_{x^{\prime }\rightarrow x}\nabla
_{i}\nabla _{k}^{\prime }\langle 0|\varphi (x)\varphi (x^{\prime })|0\rangle
+\left[ \left( \xi -\frac{1}{4}\right) g_{ik}\nabla ^{l}\nabla _{l}-\xi
\nabla _{i}\nabla _{k}\right] \langle 0|\varphi ^{2}(x)|0\rangle ,
\label{vevEMTWf}
\end{equation}%
and the expressions for the Wightman function (\ref{Wf02n}) and for VEV of
the field square. For the non-zero components one obtains
\begin{eqnarray}
\langle T_{0}^{0}\rangle _{\mathrm{ren}}^{(s)} &=&\frac{m^{D+1}}{(2\pi )^{%
\frac{D+1}{2}}}\sum_{l=1}^{q-1}\left\{ (1-4\xi )y_{l}^{2}\frac{K_{\nu
+2}(2mry_{l})}{(2mry_{l})^{\nu }}\right.   \notag \\
&&\left. +\left[ 2(4\xi -1)y_{l}^{2}-1\right] \frac{K_{\nu +1}(2mry_{l})}{%
(2mry_{l})^{\nu +1}}\right\} ,  \label{T00st} \\
\langle T_{1}^{1}\rangle _{\mathrm{ren}}^{(s)} &=&\frac{m^{D+1}}{(2\pi )^{%
\frac{D+1}{2}}}\sum_{l=1}^{q-1}\left( 4\xi y_{l}^{2}-1\right) \frac{K_{\nu
+1}(2mry_{l})}{(2mry_{l})^{\nu +1}},  \label{T11st} \\
\langle T_{2}^{2}\rangle _{\mathrm{ren}}^{(s)} &=&\frac{m^{D+1}}{(2\pi )^{%
\frac{D+1}{2}}}\sum_{l=1}^{q-1}\left( 4\xi y_{l}^{2}-1\right) \left[ \frac{%
K_{\nu +1}(2mry_{l})}{(2mry_{l})^{\nu +1}}-\frac{K_{\nu +2}(2mry_{l})}{%
(2mry_{l})^{\nu }}\right] ,  \label{T22st}
\end{eqnarray}%
and for the components in the directions parallel to the string we have (no
summation over $i$)%
\begin{equation}
\langle T_{i}^{i}\rangle _{\mathrm{ren}}^{(s)}=\langle T_{0}^{0}\rangle _{%
\mathrm{ren}}^{(s)},\;i=3,\ldots ,D.  \label{Tiist}
\end{equation}%
It can be explicitly checked that for a conformally coupled
massless scalar field this tensor is traceless. Of course, as the
background spacetime is locally flat, we could expect that the
trace anomaly is absent. From the continuity equation $\nabla
_{k}T_{i}^{k} =0$ for the energy-momentum tensor
one has the following relation for the radial and azimuthal components:%
\begin{equation}
\partial _{r}\left( r T_{1}^{1} \right) = T_{2}^{2} .
\label{conteq2}
\end{equation}%
It can be easily checked that the VEVs given by (\ref{T11st}), (\ref{T22st})
satisfy this equation. For $\xi \leqslant 1/4$ the radial stress is negative
and one has the relation $\langle T_{2}^{2}\rangle _{\mathrm{ren}%
}^{(s)}>-\nu \langle T_{1}^{1}\rangle _{\mathrm{ren}}^{(s)}$. In particular,
this is the case for minimally and conformally coupled scalar fields. In the
limit $mr\gg 1/\sin (\pi /q)$ to the leading order we have the following
relations%
\begin{eqnarray}
\langle T_{0}^{0}\rangle _{\mathrm{ren}}^{(s)} &\approx &(1-4\xi )m^{2}\sin
^{2}(\pi /q)\langle \varphi ^{2}\rangle _{\mathrm{ren}}^{(s)},\;
\label{Tii0largem} \\
\langle T_{2}^{2}\rangle _{\mathrm{ren}}^{(s)} &\approx &m^{2}\left[ 1-4\xi
\sin ^{2}(\pi /q)\right] \langle \varphi ^{2}\rangle _{\mathrm{ren}}^{(s)},
\label{T220largem}
\end{eqnarray}%
where the asymptotic expression for the VEV of the field square is given by
formula (\ref{phi20largem}). For the radial stress one has $\langle
T_{1}^{1}\rangle _{\mathrm{ren}}^{(s)}\approx -\langle T_{2}^{2}\rangle _{%
\mathrm{ren}}^{(s)}/\left[ 2mr\sin (\pi /q)\right] $. In the same limit the
energy density is positive for both minimally and conformally coupled
fields. In figure \ref{fig1} the vacuum energy density is plotted for
massive $D=3$ minimally and conformally coupled scalar fields, as a function
of $mr$ for various values of the parameter $q$. As it has been mentioned
above, for large values $mr$ the energy density tends to zero, being
positive for both minimal and conformal couplings. For the latter case the
energy density is negative near the string for $q>1$ and, hence, it has a
maximum for some intermediate value of $mr$. For a minimally coupled scalar
the energy density near the string is negative for $q^{2}>19$ and positive
for $1<q^{2}<19$.

The formulae for the components of the vacuum energy-momentum tensor in the
case of a massless scalar field are obtained in the limit $m\rightarrow 0$
(no summation over $i$):%
\begin{equation}
\langle T_{i}^{i}\rangle _{\mathrm{ren}}^{(s)}=\frac{\Gamma ((D+1)/2)}{%
2(4\pi )^{\frac{D+1}{2}}r^{D+1}}\sum_{l=1}^{q-1}\frac{f^{(i)}(y_{l})}{%
y_{l}^{D+1}},  \label{Tiisrenm0}
\end{equation}%
where%
\begin{equation}
f^{(0)}(y_{l})=(1-4\xi )(D-1)y_{l}^{2}-1,\;f^{(1)}(y_{l})=4\xi y_{l}^{2}-1,
\label{fisrenm0}
\end{equation}%
and the vacuum stresses obey the relation $\langle T_{2}^{2}\rangle _{%
\mathrm{ren}}^{(s)}=-D\langle T_{1}^{1}\rangle _{\mathrm{ren}}^{(s)}$. In
particular, for the case $D=3$, by using the expressions for $\mathcal{I}%
_{2}(0)$ and $\mathcal{I}_{4}(0)$ given before, we obtain the result derived
in \cite{Frol87,Dowk87,Smit89}. In the case of a massive scalar field, the
expression on the right of Eq. (\ref{Tiisrenm0}) determines the leading term
in the asymptotic expansion of the vacuum energy-momentum tensor near the
string.

\section{Field square and the energy-momentum tensor inside a cylindrical
shell}

\label{sec:inside}

\subsection{VEV of the field square}

\label{subsec:phi2in}

We now turn to the geometry of a string with additional
cylindrical boundary of radius $a$. Taking the coincidence limit
$x^{\prime }\rightarrow x$ in
formula (\ref{Wf2}) for the Wightman function and integrating over $\mathbf{k%
}$ with the help of the formula
\begin{equation}
\int d^{N}\mathbf{k}\int_{\sqrt{k^{2}+m^{2}}}^{\infty }\frac{k^{s}g(z)dz}{%
\sqrt{z^{2}-k^{2}-m^{2}}}=\frac{\pi ^{N/2}}{\Gamma (N/2)}B\left( \frac{N+s}{2%
},\frac{1}{2}\right) \int_{m}^{\infty }dz\,\left( z^{2}-m^{2}\right) ^{\frac{%
N+s-1}{2}}g(z),  \label{intk}
\end{equation}%
where $B(x,y)$ is the Euler beta function, the VEV of the field square is
presented as a sum of two terms%
\begin{equation}
\langle 0|\varphi ^{2}|0\rangle =\langle 0_{s}|\varphi ^{2}|0_{s}\rangle
+\langle \varphi ^{2}\rangle _{a}.  \label{phi2a}
\end{equation}%
The second term on the right of this formula is induced by the cylindrical
boundary and is given by the formula%
\begin{equation}
\langle \varphi ^{2}\rangle _{a}=-\frac{A_{D}}{\phi _{0}}\sideset{}{'}{\sum}%
_{n=0}^{\infty }\int_{m}^{\infty }dz\,z\left( z^{2}-m^{2}\right) ^{\frac{D-3%
}{2}}\frac{\bar{K}_{qn}(za)}{\bar{I}_{qn}(za)}I_{qn}^{2}(zr),  \label{phi2a1}
\end{equation}%
where we have introduced the notation%
\begin{equation}
A_{D}=\frac{2^{3-D}\pi ^{\frac{1-D}{2}}}{\Gamma \left( \frac{D-1}{2}\right) }%
.  \label{AD}
\end{equation}%
For the points away from the cylindrical surface, $r<a$, the integral in (%
\ref{phi2a1}) is exponentially convergent in the upper limit and the
boundary-induced part in the VEV of the field square is finite. In
particular this part is negative for Dirichlet scalar and is positive for
Neumann scalar. Near the string, $r\ll a$, the main contribution to $\langle
\varphi ^{2}\rangle _{a}$ comes from the summand with $n=0$ and one has%
\begin{equation}
\langle \varphi ^{2}\rangle _{a}\approx -\frac{A_{D}}{2a^{D-1}\phi _{0}}%
\int_{ma}^{\infty }dz\,\,z\left( z^{2}-m^{2}a^{2}\right) ^{\frac{D-3}{2}}%
\frac{\bar{K}_{0}(z)}{\bar{I}_{0}(z)}.  \label{phi2ar0}
\end{equation}%
As the boundary-free renormalized VEV diverges on the string, we conclude
from here that near the string the main contribution to the VEV of the field
square comes from this part. In particular, on the base of the results from
the previous section we see that for integer values $q$ the VEV of the field
square is positive near the string.

The part $\langle \varphi ^{2}\rangle _{a}$ diverges on the cylindrical
surface $r=a$. Near this surface the main contribution into (\ref{phi2a1})
comes from large values of $n$. Introducing a new integration variable $%
z\rightarrow nqz$, replacing the modified Bessel functions by their uniform
asymptotic expansions for large values of the order (see, for instance, \cite%
{hand}), and expanding over $a-r$, up to the leading order, one finds%
\begin{eqnarray}
\langle \varphi ^{2}\rangle _{a} &\approx &-\frac{(q/2a)^{D-1}(2\delta
_{B0}-1)}{\pi ^{\frac{D+1}{2}}\Gamma \left( \frac{D-1}{2}\right) }%
\int_{0}^{\infty }dz\frac{z^{D-1}}{\sqrt{1+z^{2}}}\sum_{n=1}^{\infty
}n^{D-2}e^{-2nq(1-r/a)\sqrt{1+z^{2}}}  \notag \\
&\approx &-\frac{(2\delta _{B0}-1)\Gamma \left( \frac{D-1}{2}\right) }{(4\pi
)^{\frac{D+1}{2}}(a-r)^{D-1}}.  \label{Phi2neara}
\end{eqnarray}%
This leading behavior is the same as that for a cylindrical surface of
radius $a$ in the Minkowski spacetime. As the boundary-free part is finite
at $r=a$, near the boundary the total renormalized VEV of the field square
is dominated by the boundary-induced part and is negative for Dirichlet
scalar. Combining this with the estimation for the region near the string,
we come to the conclusion that in this case the VEV\ of the field square
vanishes for some intermediate value of $r$.

Now we turn to the investigation of the boundary-induced VEV given by (\ref%
{phi2a1}), in the limiting cases of the parameter $q$. Firstly consider the
limit when the parameter $q$ is large which corresponds to small values of $%
\phi _{0}$ and, hence, to a large planar angle deficit. In this limit the
order of the modified Bessel functions for the terms with $n\neq 0$ in (\ref%
{phi2a1}) is large and we can replace these functions by their uniform
asymptotic expansions. On the base of these expansions it can be seen that
to the leading order the contribution of the terms with $n\neq 0$ is
suppressed by the factor $q^{(D-1)/2}(r/a)^{2q}$ and the main contribution
to the VEV of the field square comes from the $n=0$ term:%
\begin{equation}
\langle \varphi ^{2}\rangle _{a}\approx -\frac{A_{D}}{2\phi _{0}}%
\int_{m}^{\infty }dz\,z\left( z^{2}-m^{2}\right) ^{\frac{D-3}{2}}\frac{\bar{K%
}_{0}(za)}{\bar{I}_{0}(za)}I_{0}^{2}(zr),\;q\gg 1,  \label{phi2largeq}
\end{equation}%
with the linear dependence on $q$. In the same limit the boundary-free part
in the VEV of the field square behaves as $q^{D-1}$ and, hence, its
contribution dominates in comparison with the boundary-induced part. In the
opposite limit when $q\rightarrow 0$, the series over $n$ in Eq. (\ref%
{phi2a1}) diverges and, hence, for small values of $q$ the main contribution
comes from large values $n$. In this case, to the leading order, we can
replace the summation by the integration: $\sideset{}{'}{\sum}_{n=0}^{\infty
}f(qn)\rightarrow (1/q)\int_{0}^{\infty }dxf(x)$. As a consequence, we
obtain that in the limit $q\rightarrow 0$ the boundary-induced VEV in the
field square tends to a finite limiting value:%
\begin{equation}
\langle \varphi ^{2}\rangle _{a}=-\frac{A_{D}}{2\pi }\int_{0}^{\infty
}dx\int_{m}^{\infty }dz\,z\left( z^{2}-m^{2}\right) ^{\frac{D-3}{2}}\frac{%
\bar{K}_{x}(za)}{\bar{I}_{x}(za)}I_{x}^{2}(zr).  \label{phi2asmallq}
\end{equation}

Now we consider the limiting case obtained when $\phi _{0}\rightarrow 0$, $%
r,a\rightarrow \infty $, assuming that $a-r$ and $a\phi _{0}\equiv a_{0}$
are fixed. This corresponds to the geometry of a single boundary in the
spacetime with topology $R^{(D-1,1)}\times S^{1}$ and with $a_{0}$ being the
length for the compactified dimension. We introduce rectangular coordinates $%
(x^{\prime 1},x^{\prime 2},\ldots ,x^{\prime D})=(x,y,z_{1},\ldots ,z_{N})$
with the relations $x=a-r$, $y=a\phi $ in the limit under consideration. In
this limit, from the quantities corresponding to the geometry of a string
without a cylindrical surface we obtain the vacuum densities in the
spacetime $R^{(D-1,1)}\times S^{1}$. These quantities are well-investigated
in literature and in what follows we will consider the additional part
induced by the presence of the boundary at $x=0$. The corresponding vacuum
expectation values are obtained from the expectation values $\langle \cdots
\rangle _{a}$. For this we note that in the limit under consideration one
has $q=2\pi /\phi _{0}\rightarrow \infty $, and the order of the modified
Bessel functions in formula (\ref{phi2a1}) for $n\neq 0$ tends to infinity.
Introducing a new integration variable $z\rightarrow qnz$, we can replace
these functions by their uniform asymptotic expansions for large values of
the order. In the term with $n=0$ the arguments of the modified Bessel
functions are large and we replace these functions by the corresponding
asymptotic expressions. After straightforward calculations, the vacuum
expectation value of the field square is presented in the form%
\begin{equation}
\langle 0|\varphi ^{2}|0\rangle =\langle \varphi ^{2}\rangle ^{(0)}+\langle
\varphi ^{2}\rangle ^{(1)},  \label{phi2lim}
\end{equation}%
where $\langle \varphi ^{2}\rangle ^{(0)}$ is the vacuum expectation value
for the topology $R^{(1,D-1)}\times S^{1}$ without boundaries, and the term
\begin{equation}
\langle \varphi ^{2}\rangle ^{(1)}=-\frac{A_{D}}{2a_{0}}\sideset{}{'}{\sum}%
_{n=0}^{\infty }\int_{m_{n}}^{\infty }dz\,\left( z^{2}-m_{n}^{2}\right) ^{%
\frac{D-3}{2}}\frac{A-Bz}{A+Bz}e^{-2zx},  \label{phi2lim1}
\end{equation}%
with $m_{n}=\sqrt{m^{2}+(2\pi n/a_{0})^{2}}$, is induced by the presence of
the plate at $x=0$. In (\ref{phi2lim1}) the terms with $n\neq 0$ correspond
to the contribution of Kaluza-Klein modes related to the compactification of
the $y$ direction. The boundary induced VEV\ for the field square in the
spaces with topology $R^{(D-1,1)}\times \Sigma $ with an arbitrary internal
space $\Sigma $ is obtained in \cite{Saha06a} as a limiting case of the
corresponding braneworld geometry. It can be checked that formula (\ref%
{phi2lim1}) is a special case of this formula for $\Sigma =S^{1}$.

\subsection{VEV of the energy-momentum tensor}

\label{subsec:emtin}

The VEV of the energy-momentum tensor for the situation when the
cylindrical
boundary is present is written in the form%
\begin{equation}
\langle 0|T_{ik}|0\rangle =\langle 0_{s}|T_{ik}|0_{s}\rangle +\langle
T_{ik}\rangle _{a},  \label{Tika}
\end{equation}%
where $\langle T_{ik}\rangle _{a}$ is induced by the cylindrical boundary.
This term is obtained from the corresponding part in the Wightman function, $%
\langle \varphi (x)\varphi (x^{\prime })\rangle _{a}$, acting by the
appropriate differential operator and taking the coincidence limit [see
formula (\ref{vevEMTWf})]. For the points away from the cylindrical surface
this limit gives a finite result. For the corresponding components of the
energy-momentum tensor one obtains (no summation over $i$)%
\begin{equation}
\langle T_{i}^{i}\rangle _{a}=\frac{A_{D}}{\phi _{0}}\sideset{}{'}{\sum}%
_{n=0}^{\infty }\int_{m}^{\infty }dzz^{3}\left( z^{2}-m^{2}\right) ^{\frac{%
D-3}{2}}\frac{\bar{K}_{qn}(za)}{\bar{I}_{qn}(za)}F_{qn}^{(i)}\left[
I_{qn}(zr)\right] ,  \label{Tiia21}
\end{equation}%
with the notations%
\begin{eqnarray}
F_{qn}^{(0)}\left[ f(y)\right]  &=&\left( 2\xi -\frac{1}{2}\right) \left[
f^{\prime 2}(y)+\left( 1+\frac{q^{2}n^{2}}{y^{2}}\right) f^{2}(y)\right] +%
\frac{y^{2}-m^{2}r^{2}}{(D-1)y^{2}}f^{2}(y),  \label{ajpm} \\
F_{qn}^{(1)}\left[ f(y)\right]  &=&\frac{1}{2}f^{\prime 2}(y)+\frac{2\xi }{y}%
f(y)f^{\prime }(y)-\frac{1}{2}\left( 1+\frac{q^{2}n^{2}}{y^{2}}\right)
f^{2}(y),  \label{ajpm1} \\
F_{qn}^{(2)}\left[ f(y)\right]  &=&\left( 2\xi -\frac{1}{2}\right) \left[
f^{\prime 2}(y)+\left( 1+\frac{q^{2}n^{2}}{y^{2}}\right) f^{2}(y)\right] +%
\frac{q^{2}n^{2}}{y^{2}}f^{2}(y)-\frac{2\xi }{y}f(y)f^{\prime }(y),
\label{ajpm2}
\end{eqnarray}%
and
\begin{equation}
F_{qn}^{(i)}\left[ f(y)\right] =F_{qn}^{(0)}\left[ f(y)\right] ,\;i=3,\ldots
,D.  \label{Fi0}
\end{equation}%
It can be checked that the expectation values (\ref{Tiia21}) satisfy
equation (\ref{conteq2}) and, hence, the continuity equation for the
energy-momentum tensor. The boundary-induced part in the VEV of the
energy-momentum tensor given by Eq. (\ref{Tiia21}) is finite everywhere
except at the points on the boundary and at the points on the string in the
case $q<1$. As we will see below, unlike to the surface divergences, the
divergences on the string are integrable.

In the case $q>1$, near the string, $r\rightarrow 0$, the main contribution
to the boundary part (\ref{Tiia21}) comes from the summand with $n=0$ and
one has
\begin{equation}
\langle T_{i}^{i}\rangle _{a}\approx \frac{A_{D}}{2\phi _{0}a^{D+1}}%
\int_{ma}^{\infty }dzz^{3}\left( z^{2}-m^{2}a^{2}\right) ^{\frac{D-3}{2}}%
\frac{\bar{K}_{0}(z)}{\bar{I}_{0}(z)}F^{(i)}(z)  \label{Tii21r0}
\end{equation}%
with the notations%
\begin{equation}
F^{(0)}(z)=2\xi -\frac{1}{2}+\frac{1-m^{2}a^{2}/z^{2}}{D-1},\;F^{(i)}(z)=\xi
-\frac{1}{2},\;i=1,2.  \label{F0z}
\end{equation}%
For $q<1$ the main contribution to the boundary-induced part for the points
near the string comes from $n=1$ term and in the leading order one has%
\begin{equation}
\langle T_{i}^{i}\rangle _{a}\approx \frac{q^{2}A_{D}r^{2q-2}F_{1}^{(i)}}{%
2^{q}\pi \Gamma ^{2}(q+1)}\int_{m}^{\infty }dzz^{2q+1}\left(
z^{2}-m^{2}\right) ^{\frac{D-3}{2}}\frac{\bar{K}_{q}(za)}{\bar{I}_{q}(za)},
\label{Tii21r0q}
\end{equation}%
where%
\begin{equation}
F_{1}^{(0)}=q(2\xi -1/2),\;F_{1}^{(1)}=\xi ,\;F_{1}^{(2)}=(2q-1)\xi .
\label{Fi1}
\end{equation}%
As we see, in this case the VEVs for the energy-momentum tensor diverge on
the string. This divergence is integrable. In particular, the corresponding
contribution to the energy in the region near the string is finite.

\begin{figure}[tbph]
\begin{center}
\epsfig{figure=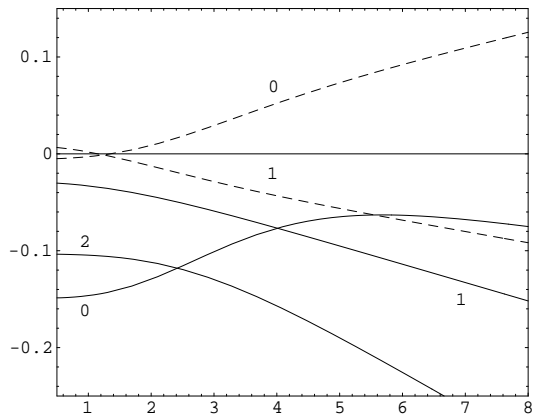,width=7.cm,height=5.5cm}
\end{center}
\caption{Boundary-induced parts in the components of the energy-momentum
tensor, $a^{D+1}\langle T_{i}^{i}\rangle _{a}$, evaluated at $r/a=0.5$ as
functions of the parameter $q$ for minimally (full curves) and conformally
(dashed curves) coupled $D=3$ massless scalar fields with Dirichlet boundary
condition. The numbers near the curves correspond to the value of the index $%
i$. The azimuthal stress for the case of conformally coupled field is
obtained from the zero-trace condition. }
\label{fig2}
\end{figure}

As in the case of the field square, in the limit $q\gg 1$ the contribution
of the terms with $n\neq 0$ to the VEV of the energy-momentum tensor is
suppressed by the factor $q^{(D-1)/2}(r/a)^{2q}$ and the main contribution
comes from the $n=0$ term with the linear dependence on $q$. In the same
limit, the boundary-free part in the VEV\ of the energy-momentum tensor
behaves as $q^{D+1}$ and, hence, the total energy-momentum tensor is
dominated by this part. In the opposite limit when $q\ll 1$, by the way
similar to that used before for the VEV of the field square, it can be seen
that the boundary-induced part in the vacuum energy-momentum tensor tends to
a finite limiting value which is obtained from (\ref{Tiia21}) replacing the
summation over $n$ by the integration. The described behavior of the VEVs as
functions of $q$ is clearly seen in figure \ref{fig2} where the dependence
of the boundary-induced vacuum energy density and stresses on the parameter $%
q$ is plotted in the case of $D=3$ minimally and conformally coupled
massless scalar fields with Dirichlet boundary condition and for $r/a=0.5$.

The boundary part $\left\langle T_{i}^{k}\right\rangle _{a}$ diverges on the
cylindrical surface $r=a$. Introducing a new integration variable $%
z\rightarrow nqz$ and taking into account that near the surface $r=a$ the
main contribution comes from large values of $n$, we can replace the
modified Bessel functions by their uniform asymptotic expansions for large
values of the order. To the leading order this gives%
\begin{equation}
\langle T_{i}^{i}\rangle _{a}\approx \frac{D(\xi -\xi _{D})(2\delta _{B0}-1)%
}{2^{D}\pi ^{(D+1)/2}(a-r)^{D+1}}\Gamma \left( \frac{D+1}{2}\right) ,\quad
i=0,2,\ldots ,D.  \label{T00asra2}
\end{equation}%
This leading divergence does not depend on the parameter $q$ and coincides
with the corresponding one for a cylindrical surface of radius $a$ in the
Minkowski bulk. For the radial component to the leading order one has $%
\langle T_{1}^{1}\rangle _{a}\sim (a-r)^{-D}$. In particular, for a
minimally coupled scalar field the corresponding energy density is negative
for Dirichlet boundary condition and is positive for non-Dirichlet boundary
conditions. For a conformally coupled scalar the leading term vanishes and
it is necessary to keep the next term in the corresponding asymptotic
expansion. As the boundary-free part in the VEV of the energy-momentum
tensor is finite on the cylindrical surface, for the points near the
boundary the vacuum energy-momentum tensor is dominated by the
boundary-induced part.

In the limit $\phi _{0}\rightarrow 0$, $r,a\rightarrow \infty $, with fixed
values for $a-r$ and $a\phi _{0}\equiv a_{0}$, proceeding in a similar way
to that used for the field square, the VEV of the energy-momentum tensor can
be written as
\begin{equation}
\langle 0|T_{i}^{i}|0\rangle =\langle T_{i}^{i}\rangle ^{(0)}+\langle
T_{i}^{i}\rangle ^{(1)},  \label{Tiilimn}
\end{equation}%
where $\langle T_{i}^{i}\rangle ^{(0)}$ is the corresponding quantity for
the topology $R^{(1,D-1)}\times S^{1}$ without boundaries, and the term (no
summation over $i$)%
\begin{equation}
\langle T_{i}^{i}\rangle ^{(1)}=\frac{A_{D}}{2a_{0}}\sideset{}{'}{\sum}%
_{n=0}^{\infty }\int_{m_{n}}^{\infty }dz\,\left( z^{2}-m_{n}^{2}\right) ^{%
\frac{D-3}{2}}\frac{A-Bz}{A+Bz}F_{1}^{(i)}(z)e^{-2zx},  \label{TiilimnBi}
\end{equation}%
is induced by the boundary located at $x=0$. In (\ref{TiilimnBi}), $m_{n}$
is defined by the relation given in the paragraph after formula (\ref%
{phi2lim1}) and we have introduced the notations%
\begin{eqnarray}
F_{1}^{(0)}(z) &=&(4\xi -1)z^{2}+\frac{z^{2}-m_{n}^{2}}{D-1},  \label{F10z0}
\\
F_{1}^{(2)}(z) &=&(4\xi -1)z^{2}+\left( \frac{2\pi n}{a_{0}}\right) ^{2},
\label{F10z2}
\end{eqnarray}%
with $F_{1}^{(1)}(z)=0$. In particular, in this limit the boundary-induced
vacuum stress in the direction perpendicular to the plate vanishes. Formula (%
\ref{TiilimnBi}) is a special case of the more general result for the spaces
with topology $R^{(D-1,1)}\times \Sigma $ with an arbitrary internal space $%
\Sigma $ obtained in \cite{Saha06b} as a limiting case of the corresponding
braneworld geometry.

\begin{figure}[tbph]
\begin{center}
\begin{tabular}{cc}
\epsfig{figure=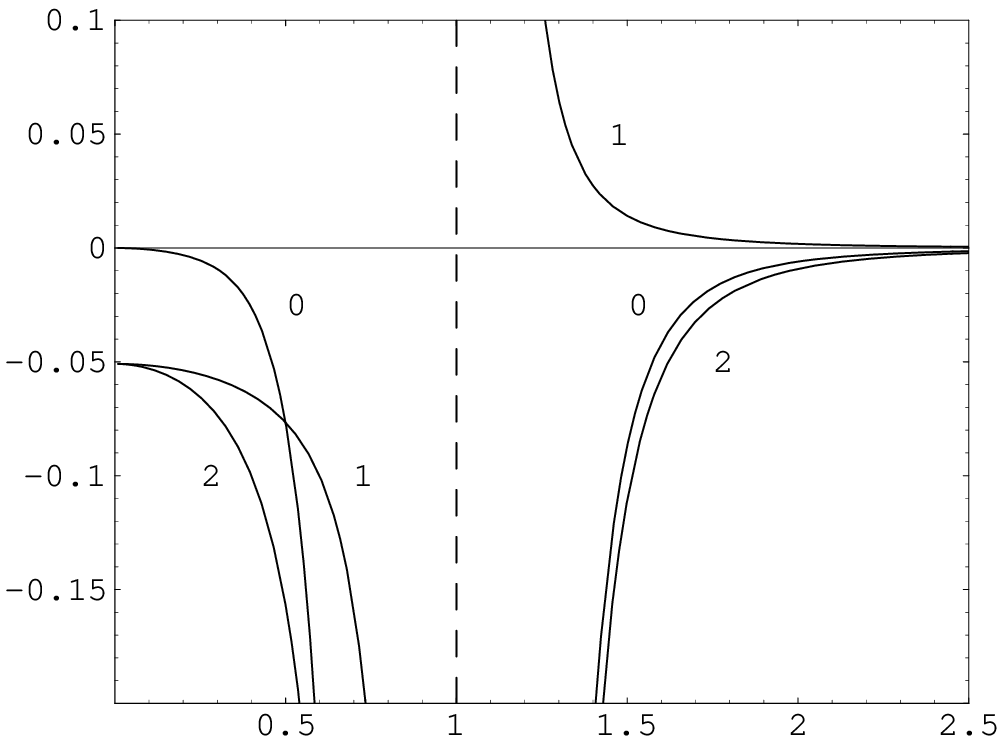,width=7.cm,height=5.5cm} & \quad %
\epsfig{figure=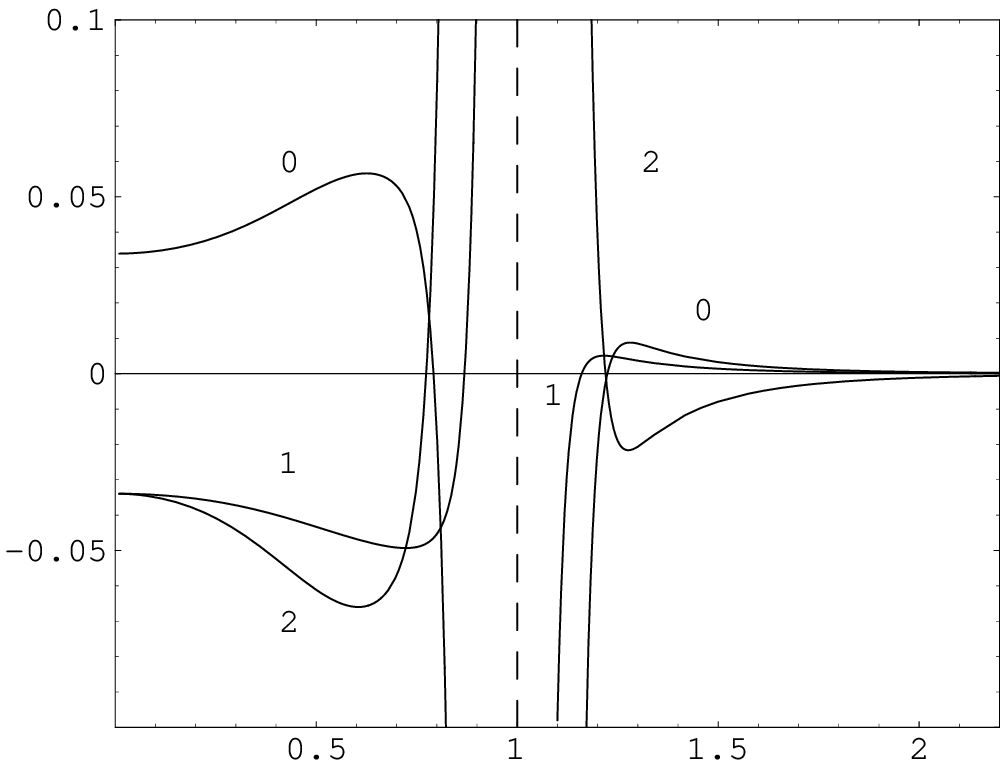,width=7.cm,height=5.5cm}%
\end{tabular}%
\end{center}
\caption{Boundary-induced parts of the components of the energy-momentum
tensor multiplied by $a^{D+1}$, $a^{D+1}\langle T_{i}^{i}\rangle _{a}$, as
functions of $r/a$ in both interior and exterior regions of the cylindrical
surface. The graphs are plotted for minimally (left panel) and conformally
(right panel) coupled $D=3$ massless scalar fields with Dirichlet boundary
condition on the cylindrical surface and for the cosmic string bulk with $q=4
$. The numbers near the curves correspond to the value of the index $i$.}
\label{fig3}
\end{figure}

In figure \ref{fig3} we present the graphs for the boundary induced parts of
the components of the energy-momentum tensor as functions of $r/a$. The
graphs are plotted for $D=3$ massless scalar field with Dirichlet boundary
condition on the cylindrical surface and for the cosmic string bulk with $q=4
$. In this special case for a minimally coupled scalar field, the
boundary-induced vacuum energy density vanishes on the cosmic string.

\section{VEVs in the region outside a cylindrical shell}

\label{sec:outside}

In this section we consider the VEVs induced by the cylindrical boundary in
the exterior region $r>a$. Taking the coincidence limit for the arguments,
from the corresponding formula for the Wightman function we obtain the
boundary-induced part in the VEV of the field square:
\begin{equation}
\langle \varphi ^{2}\rangle _{a}=-\frac{A_{D}}{\phi _{0}}\sideset{}{'}{\sum}%
_{n=0}^{\infty }\int_{m}^{\infty }dz\,z\left( z^{2}-m^{2}\right) ^{\frac{D-3%
}{2}}\frac{\bar{I}_{qn}(za)}{\bar{K}_{qn}(za)}K_{qn}^{2}(zr),
\label{phi2a1ext}
\end{equation}%
with $A_{D}$ defined by Eq. (\ref{AD}). As for the interior region, the
expression on the right diverges on the cylindrical surface. The leading
term in the corresponding asymptotic expansion near this surface is obtained
from that for the interior region, formula (\ref{Phi2neara}), replacing $%
(a-r)$ by $(r-a)$. For large distances from the cylindrical surface, $r\gg a$%
, and for a massless scalar field, we introduce in (\ref{phi2a1ext}) a new
integration variable $y=zr/a$ and expand the integrand over $a/r$. The main
contribution comes from the $n=0$ term. By taking into account the value for
the standard integral involving the square of the MacDonald function \cite%
{Prud86}, to the leading order for $A\neq 0$, one finds
\begin{equation}
\langle \varphi ^{2}\rangle _{a}\approx -\frac{\pi ^{1-D/2}}{2^{D}\phi
_{0}r^{D-1}\ln (r/a)}\frac{\Gamma ^{2}\left( \frac{D-1}{2}\right) }{\Gamma
(D/2)}.  \label{phi2larger}
\end{equation}%
In the case of Neumann boundary condition ($A=0$) and $q>1$, the leading
contribution again comes from the $n=0$ term:%
\begin{equation}
\langle \varphi ^{2}\rangle _{a}\approx \frac{\pi ^{1-D/2}(D-1)}{\phi
_{0}a^{D-1}}\frac{\Gamma ^{2}\left( \frac{D+1}{2}\right) }{\Gamma (D/2)}%
\left( \frac{a}{2r}\right) ^{D+1}.  \label{phi2largerNeu}
\end{equation}%
For Neumann boundary condition and $q<1$, at large distances the main
contribution comes from the $n=1$ term and the VEV of the field square
behaves as $1/r^{D+2q-1}$. As we see, the boundary-free part in the VEV
dominates at large distances from the boundary. For a massive field, under
the condition $mr\gg 1$ the main contribution into the integral in (\ref%
{phi2a1ext}) comes from the lower limit and one obtains%
\begin{equation}
\langle \varphi ^{2}\rangle _{a}\approx -\frac{\pi m^{\frac{D-3}{2}}e^{-2mr}%
}{(4\pi )^{\frac{D-1}{2}}\phi _{0}r^{\frac{D+1}{2}}}\sideset{}{'}{\sum}%
_{n=0}^{\infty }\frac{\bar{I}_{qn}(ma)}{\bar{K}_{qn}(ma)},
\label{phi2largerm}
\end{equation}%
with the exponential suppression of the boundary induced VEV. Note that here
the suppression is stronger compared with the boundary-free part. For large
values $q\gg 1$, the contribution of the terms with $n\neq 0$ to the VEV (%
\ref{phi2a1ext}) can be estimated by using the uniform asymptotic expansions
for the modified Bessel functions when the order is large. This contribution
is suppressed by the factor $q^{(D-1)/2}(a/r)^{2q}$ and the main
contribution to the VEV of the field square comes from the $n=0$ term which
is a linear function on $q$. In the opposite limit, $q\ll 1$, analogously to
the procedure for the interior region, it can be seen that $\langle \varphi
^{2}\rangle _{a}$ tends to the finite value which is obtained from (\ref%
{phi2a1ext}) replacing the summation over $n$ by the integration.

For the part in the vacuum energy-momentum tensor induced by the cylindrical
surface in the region $r>a$, from (\ref{Wfa0ext}), (\ref{vevEMTWf}), (\ref%
{phi2a1ext}) one has the following formula%
\begin{equation}
\langle T_{i}^{i}\rangle _{a}=\frac{A_{D}}{\phi _{0}}\sideset{}{'}{\sum}%
_{n=0}^{\infty }\int_{m}^{\infty }dzz^{3}\left( z^{2}-m^{2}\right) ^{\frac{%
D-3}{2}}\frac{\bar{I}_{qn}(za)}{\bar{K}_{qn}(za)}F_{qn}^{(i)}\left[
K_{qn}(zr)\right] ,  \label{Tiiext}
\end{equation}%
with the functions $F_{qn}^{(i)}\left[ f(y)\right] $ defined by formulae (%
\ref{ajpm})-(\ref{ajpm2}). The VEVs given by formula (\ref{Tiiext}) diverge
at the points on the bounding surface with the leading divergence obtained
from analog formula in the interior region, Eq. (\ref{T00asra2}), by the
replacement $a-r\rightarrow r-a$. By the way similar to that used above for
the vacuum expectation value of the field square, it can be seen that for a
massless scalar field with $A\neq 0$ and at large distances from the
cylindrical surface, $r\gg a$, the components of the vacuum energy-momentum
tensor behave as%
\begin{equation}
\langle T_{0}^{0}\rangle _{a}\approx -(D-1)\langle T_{1}^{1}\rangle
_{a}\approx \frac{D-1}{D}\langle T_{2}^{2}\rangle _{a}\approx \frac{\Gamma
^{2}\left( \frac{D+1}{2}\right) }{(4\pi )^{D/2-1}\Gamma (D/2)}\frac{\xi -\xi
_{D}}{\phi _{0}r^{D+1}\ln (r/a)}.  \label{Tiilarger}
\end{equation}%
For a conformally coupled scalar field this leading terms vanish and it is
necessary to take into account the next terms in the asymptotic expansion.
In the case of Neumann boundary condition the components of the
energy-momentum tensor behave as $1/r^{D+3}$ when $q>1$ and as $1/r^{D+2q+1}$
when $q<1$. For a massive scalar field the main contribution comes from the
lower limit of the integral in Eq. (\ref{Tiiext}) and we find%
\begin{equation}
\langle T_{0}^{0}\rangle _{a}\approx -2mr\langle T_{1}^{1}\rangle
_{a}\approx \langle T_{2}^{2}\rangle _{a}\approx \frac{\left( 4\xi -1\right)
e^{-2mr}}{2^{D-1}\pi ^{(D-3)/2}}\left( \frac{m}{r}\right) ^{\frac{D+1}{2}}%
\sideset{}{'}{\sum}_{n=0}^{\infty }\frac{\bar{I}_{qn}(ma)}{\bar{K}_{qn}(ma)}.
\label{Tiilargerm}
\end{equation}%
As we see in this case the radial stress is suppressed compared with the
other components by an additional factor $mr$. As in the case of the field
square, for large values of the parameter $q\gg 1$ the main contribution to
the VEV of the energy-momentum tensor comes from the $n=0$ term which is a
linear function on $q$. For small values $q\ll 1$, in the leading order the
summation over $n$ can be replaced by the integration and the VEV tends to
the finite limiting value. In figure \ref{fig3} we have plotted the
dependence of the boundary-induced parts in the components of the vacuum
energy-momentum tensor for the exterior region as functions on $r/a$ in the
case of $D=3$ Dirichlet massless scalar with minimal and conformal couplings.

In the discussion above we have considered the idealized geometry of a
cosmic string with zero thickness. A realistic cosmic string has a structure
on a length scale defined by the phase transition at which it is formed. As
it has been shown in Refs. \cite{Alle90,Alle96}, for a non-minimally coupled
scalar field the internal structure of the string has non-negligible effects
even at large distances. Note that when the cylindrical boundary is present
with the boundary condition (\ref{Dirbc}), the VEVs of the physical
quantities in the exterior region are uniquely defined by the boundary
conditions and the bulk geometry. This means that if we consider a
non-trivial core model with finite thickness $b<a$ and with the line element
(\ref{ds21}) in the region $r>b$, the results in the region outside the
cylindrical shell will not be changed. As regards to the interior region,
the formulae given above are the first stage of the evaluation of the VEVs
and other effects could be present in a realistic cosmic string.

\section{VEVs for a cosmic string with finite thickness}

\label{sec:finitethick}

From the point of view of the physics in the exterior region the cylindrical
surface with boundary condition (\ref{Dirbc}) can be considered as a simple
model of cosmic string core. In general, the string core is modelled by a
cylindrically symmetric potential whose support lies in $r\leqslant a$. In
this section we generalize the results in the exterior region for the model
of core described by the line-element \cite{Alle90,Alle96}%
\begin{equation}
ds^{2}=dt^{2}-P^{2}(r/a)dr^{2}-r^{2}d\phi ^{2}-\sum_{i=1}^{N}dz_{i}{}^{2},
\label{ds2f}
\end{equation}%
where $P(x)$ is a smooth monotonic function satisfying the conditions%
\begin{equation}
\lim_{x\rightarrow 0}P(x)=1/q,\;P(x)=1\;\mathrm{for}\;x>1.  \label{condP}
\end{equation}%
The eigenfunctions in the region $r<a$ have the structure (\ref{eigfunccirc}%
) with the radial function $f_{n}(r/a,\gamma a)$ instead of $\beta _{\alpha
}J_{q\left\vert n\right\vert }(\gamma r)$. The equation for the radial
function is obtained from field equation (\ref{fieldeq}) with the metric
given by (\ref{ds2f}):%
\begin{equation}
\left[ \frac{1}{xP(x)}\frac{d}{dx}\frac{x}{P(x)}\frac{d}{dx}+\gamma
^{2}a^{2}-\frac{q^{2}n^{2}}{x^{2}}-\frac{2\xi }{x}\frac{P^{\prime }(x)}{%
P^{3}(x)}\right] f_{n}(x,\gamma a)=0.  \label{Rneq}
\end{equation}%
We will denote by $R_{n}(x,\gamma a)$ the solution of this equation regular
at $r=0$. Now the radial part of the eigenfunctions is written in the form%
\begin{equation}
\begin{array}{ll}
R_{n}(r/a,\gamma a) & \mathrm{for}\;r<a \\
A_{n}J_{q\left\vert n\right\vert }(\gamma r)+B_{n}Y_{q\left\vert
n\right\vert }(\gamma r)\; & \mathrm{for}\;r>a.%
\end{array}
\label{eigfuncf}
\end{equation}%
The coefficients $A_{n}$ and $B_{n}$ are determined from the conditions of
the continuity of the radial functions and their derivatives at $r=a$:%
\begin{eqnarray}
A_{n} &=&\frac{\pi }{2}\left[ \gamma aY_{q\left\vert n\right\vert }^{\prime
}(\gamma a)R_{n}(1,\gamma a)-Y_{q\left\vert n\right\vert }(\gamma
a)R_{n}^{\prime }(1,\gamma a)\right] ,  \label{Anf} \\
B_{n} &=&-\frac{\pi }{2}\left[ \gamma aJ_{q\left\vert n\right\vert }^{\prime
}(\gamma a)R_{n}(1,\gamma a)-J_{q\left\vert n\right\vert }(\gamma
a)R_{n}^{\prime }(1,\gamma a)\right] ,  \label{Bnf}
\end{eqnarray}%
where $R_{n}^{\prime }(1,\gamma a)=(d/dx)R_{n}(x,\gamma a)|_{x=1}$.
Substituting these expressions into the formula for the radial
eigenfunctions in the region $r>a$, we see that these eigenfunctions have
the form (\ref{replace}) where the barred notations are obtained from from
the expressions given by Eq. (\ref{fbar}) by the replacement $A/B\rightarrow
-R_{n}^{\prime }(1,\gamma a)/R_{n}(1,\gamma a)$. The further evaluation of
the Wightman function in the region $r>a$ is similar to that described in
subsection \ref{subsec:exterior}. Therefore, the part in the Wightman
function induced by the non-trivial structure of the string core is given by
formula (\ref{Wfa0ext}), where in the expressions for the definition of the
barred modified Bessel functions [see Eq. (\ref{fbar})] we should substitute%
\begin{equation}
\frac{A}{B}=-\frac{R_{n}^{\prime }(1,zae^{\pi i/2})}{R_{n}(1,zae^{\pi i/2})}.
\label{ABreplacef}
\end{equation}%
The formulae for the VEVs of the field square and the energy-momentum tensor
are obtained from formulae (\ref{phi2a1ext}) and (\ref{Tiiext}) by the same
substitution. The corresponding results for the Euclidean Green function and
the VEV of the field square in a special case of $D=3$ massless scalar field
are given in Ref. \cite{Alle96}. Two specific models of the string core have
been considered in literature. In the "ballpoint pen" model \cite{Hisc85}
the region $r<a$ has a constant curvature, whereas in the "flower pot" model
\cite{Alle90} the curvature of spacetime is concentrated on a ring of radius
$a$. In these models the Euclidean Green function and the VEV of the field
square for $D=3$ massless scalar field are investigated in Ref. \cite{Alle90}%
. For the first model one has%
\begin{equation}
\frac{A}{B}=-\sqrt{q^{2}-1}\frac{P_{\nu }^{|n|\prime }(1/q)}{P_{\nu
}^{|n|}(1/q)},\;\nu (\nu +1)=-2\xi -\frac{z^{2}a^{2}}{q^{2}-1}.
\label{ballpoint}
\end{equation}%
where $P_{\nu }^{|n|}(x)$ is the associated Legendre function. For the
second model the curvature is a delta function concentrated on a ring of
radius $a$ and the radial parts of the eigenfunctions have a discontinuity
in their slope at $r=a$. The corresponding jump condition is obtained by
integrating the radial part of the field equation through the point $r=a$.
This procedure leads to the ratio of the coefficients given by the formula
\begin{equation}
\frac{A}{B}=-za\frac{I_{|n|}^{\prime }(za/q)}{I_{|n|}(za/q)}-2\xi (q-1).
\label{flowerpot}
\end{equation}%
For the first model in (2+1)-dimensions the ground state energy of a massive
scalar field is investigated in Ref. \cite{Khus99}.

\section{Conclusion}

\label{sec:Conc}

We have investigated the local one-loop quantum effects for a massive scalar
field induced by a cylindrical boundary in the spacetime of a cosmic string.
We have assumed that on the bounding surface the field obeys Robin boundary
condition. The latter is a generalization of Dirichlet and Neumann boundary
conditions and arises in a variety of physical situations. As a first step
in the evaluation of the renormalized VEVs of the field square and the
energy-momentum tensor, in section \ref{sec:WightFunc} we have considered
the Wightman function in both interior and exterior regions. The
corresponding mode-sum in the interior region contains the summation over
the zeros of a combination of the Bessel function and its derivative. For
the summation of the corresponding series we have used the generalized
Abel-Plana formula which allows us to extract from the mode sum the Wightman
function for the cosmic string background without the cylindrical shell and
to present the boundary-induced part in terms of exponentially convergent
integrals in the coincidence limit for the points away from the boundary.
The representation of the Wightman function where the boundary-free part is
explicitly extracted is given also for the exterior region. The
boundary-induced parts in the interior and exterior Wightman functions are
related by the replacements $I_{qn}\rightleftarrows K_{qn}$.

In section \ref{sec:noboundary} we have considered the VEVs induced by the
cosmic string geometry without boundaries. Though this geometry is
well-investigated in literature, to our knowledge, no closed formulae were
given for the VEVs of the field square and the energy-momentum tensor for a
massive field with general curvature coupling in an arbitrary number of
dimensions. We show that such formulae can be derived when the parameter $q$
is an integer. In this case the corresponding Wightman function is the image
sum of the Minkowskian Wightman functions. Renormalized VEVs of the field
square and the energy-momentum tensor are determined by formulae (\ref{phi2w}%
), (\ref{T00st})-(\ref{T22st}) for a massive field and by formulae (\ref%
{phi2wmassless}), (\ref{Tiisrenm0}) for a massless one. In the latter case
and for odd values of the spatial dimension the summation over $l$ can be
done by using the recurrent formula. In this case the VEVs are polynomial
functions of $q$ and by analytic continuation the corresponding formulae are
valid for all values of this parameter. By using the formula for the
interior Wightman function, in section \ref{sec:inside} we have investigated
the VEVs of the field square and the energy-momentum tensor in this region.
The corresponding boundary-induced parts are given by formulae (\ref{phi2a1}%
) and (\ref{Tiia21}). For the points on the string these parts are finite
when $q\geqslant 1$. For $q<1$ the boundary-induced part in the VEV of the
field square remains finite on the string, but the corresponding part in the
energy-momentum tensor has integrable divergences. Near the string the
boundary-free parts behave as $1/r^{D-1}$ for the field square and as $%
1/r^{D+1}$ for the energy-momentum tensor and these parts dominate. For the
points near the boundary the situation is opposite and the boundary-induced
parts are dominant. For large values of the parameter $q$ the
boundary-induced VEVs for both field square and the energy-momentum tensor
are linear functions of this parameter, whereas for small values of $q$ they
tend to finite limiting value. We have the similar behavior for the VEVs in
the region outside the cylindrical shell. These VEVs are investigated in
section \ref{sec:outside} and are given by formulae (\ref{phi2a1ext}) and (%
\ref{Tiiext}). In the case of a massless field with non-Neumann boundary
condition, at large distances from the cylindrical surface the
boundary-induced parts behave as $1/\left[ r^{D-1}\ln (r/a)\right] $ for the
field square and as $1/\left[ r^{D+1}\ln (r/a)\right] $ for the
energy-momentum tensor. For a massive field under the assumption $mr\gg 1$,
the boundary induced VEVs are exponentially suppressed. This suppression is
stronger than that for the boundary-free parts.

The cylindrical surface with boundary condition (\ref{Dirbc}) can be
considered as a simple model of the cosmic string core. In section \ref%
{sec:finitethick} we give the generalization of the corresponding results in
the exterior region for a general cylindrically symmetric static model of
the string core with finite support. We have shown that the corresponding
formulae are obtained from the formulae for the cylindrical surface with the
substitution given by Eq. (\ref{ABreplacef}), where $R_{n}(r/a,\gamma a)$ is
a regular solution of the radial equation for the eigenfunctions. For two
special models of the string core, namely, the "ballpoint pen" and "flower
pot" models, the ratio of the coefficients is given by formulae (\ref%
{ballpoint}) and (\ref{flowerpot}), respectively.

\section*{Acknowledgments}

AAS was supported by PVE/CAPES Program and in part by the Armenian Ministry
of Education and Science Grant No. 0124. ERBM and VBB thank Conselho
Nacional de Desenvolvimento Cient\'{\i}fico e Tecnol\'{o}gico (CNPq) and
FAPESQ-PB/CNPq (PRONEX) for partial financial support.

\end{document}